\documentclass[runningheads]{llncs}
\usepackage{graphicx}
\usepackage{amsmath}
\usepackage{amssymb}
\usepackage{bm}
\usepackage{multirow}

\def\FF{\mathbb{F}}
\def\ZZ{\mathbb{Z}}

\def\mincut{\mathop{\rm mincut}\nolimits}

\def\Label#1{\label{#1}\ [\ \text{#1}\ ]\ }
\def\Label{\label}

\begin{document}
\title{Secure network coding with adaptive and active  attack\thanks{Supported 
in part by the National Natural Science Foundation of China
under Grant 62171212.}}
%
%
\author{Masahito~Hayashi\inst{1,2,3}\orcidID{0000-0003-3104-1000}}
\authorrunning{M. Hayashi}
%
\institute{School of Data Science, The Chinese University of Hong Kong, Shenzhen, Longgang
518172, China \and
International Quantum Academy (SIQA),
Futian, Shenzhen 518048, China \and
Graduate School
of Mathematics, Nagoya University, Chikusa-ku, Nagoya 464-8602, Japan\\
\email{hmasahito@cuhk.edu.cn}}
\maketitle              
\begin{abstract}
Ning Cai and the author jointly studied 
secure network codes over adaptive and active attacks, 
which were rarely studied until these seminal papers.
This paper reviews the result for secure network code over adaptive and active attacks.
We focus on two typical network models, a one-hop relay network and a unicast relay network.
\keywords{one-hop relay network  \and unicast relay network \and 
non-linear code \and
anti-Latin square.}
\end{abstract}
\section{Introduction}\Label{S1}
Network coding is a key technology for communication over a network.
Secure network coding protects our communication 
from various attacks.
However, the majority of existing studies
addressed deterministic and passive attacks, and 
did not care about the effect of adaptive changes on attacked edges
and active modification on the information on attacked edge \cite{Cai2002,CY,YN,CG,RSS,FMSS,NYZ,HY,Cai,CC,SK,KMU,Matsumoto2011,Matsumoto2011a}. 
The communication over a network has a risk that 
an adversary, Eve, can improve her ability by using 
adaptive and active operations.
In several joint papers 
\cite{HOKC17,HC21a,HC21b,HOKC,CH},
Ning Cai and the author jointly studied 
how adaptive and active operations can improve Eve's performance. 
This paper reviews these joint studies.
Since the original paper \cite{CH} contains complicated descriptions
and a certain error,
this paper presents the corrected contents based on its correction \cite{CH-cor}.
In addition, the paper \cite{CH}
discussed the difference between the existence and the non-existence
of random number at intermediate nodes.
Also, the paper \cite{CH} treated the correctness and the secrecy
in an asymmetric way.
In this paper, we adopt the base $2$ of the logarithm.

The outline of this paper is the following.
This paper explains why such an asymmetric treatment is reasonable
by citing the recent paper \cite{WLH} in Section \ref{S2}.
Then, Section \ref{S2} introduces these two kinds of formulations
based on the existence and the non-existence
of random number at intermediate nodes.
Section \ref{S9} analyzes $r$-wiretap network
based on the above mentioned formulations.
Section \ref{SS} introduces two kinds of unconventional attack models,
adaptive attack and active attack. 
Section \ref{S3} explains that
secure communication is impossible in
one-hop relay network 
under the non-existence
of random number at intermediate nodes even when 
neither adaptive attack nor active attack is allowed for Eve.
Section \ref{S4} introduces non-linear codes
to resolve this problem.
Section \ref{S5} shows adaptive attacks the above non-linear codes
over one-hop relay network.
Section \ref{S6} shows active attacks the above non-linear codes
over one-hop relay network.
Section \ref{S7} presents the usage of vector-linear code 
over one-hop relay network.
Section \ref{S8} 
derives the capacities 
under adaptive and active attacks
over unicast relay network.
Section \ref{S11} makes conclusion. 

\section{Security model and coding model}\Label{S2}
Secure network coding 
has two kinds of security measures.
One is the correctness of the decoding message,
and the other is the secrecy of the transmitted message.
Also, secure network coding considers the set of possible attacks.
The usual setting requires the correctness and the secrecy
when one of possible attack is done.
However, when this requirement is imposed to our code,
it is often difficult to realize a larger transmission rate.
To realize a better transmission rate, 
it is possible to discount the requirement as follows.
The discounted setting requires
the secrecy when one of possible attack is done,
but it requires the correctness only when no attack is made.
This requirement is useful 
when the sender, Alice, and the receiver, Bob, share
a small size of secret random number and they can use public channel
due to the following reason.

As explained in \cite{WLH}, 
this discounted setting is reasonable as follows
even for the message transmission.
In this case, after the communication,
Alice and Bob can apply {\it error verification}
by using a small size of secret random number and public channel
\cite[Section VI]{WLH}, \cite[Section VIII]{FMC}, \cite[Section III-C-2)]{H22}.
If an error is detected, they abort the protocol.
Since the attack is limited to an element of the set of possible attacks,
the requirement guarantees that the secrecy of the message is kept
even in this case.
Once the error verification is passed,
the error verification guarantees the correctness.
As explained in 
\cite[Section VI]{WLH}, \cite[Section VIII]{FMC}, \cite[Section III-C-2)]{H22},
when the size of the transmitted message is $n$ bits,
the error verification consumes a polynomial amount of resources
with respect to $\log n$.
Thus, the error verification does not affect the transmission rate of the whole protocol.
Due to this reason,
this paper focuses on the discounted setting.

This paper adopts two kinds of formulations for 
our codes depending on the resources on intermediate nodes.
In the first formulation,
an arbitrary node including an intermediate node
can generate an arbitrary random number.
In the second formulation,
only the source node and the destination node
can generate an arbitrary random number so that an intermediate node cannot.
This paper adopts both formulations for our codes.

This paper considers two kinds of linear codes.
When transmitted information on each edge in a code
is given as a single element, i.e., a scalar, of
a finite field $\FF_q$ or a quotient ring $\ZZ_d$
and 
a coding operation on each node is given as linear operations
over input or incoming information, 
the code is called a {\it scalar-linear code}.
 When transmitted information on each edge in a code
is given as a vector on a finite field $\FF_q$ or a quotient ring $\ZZ_d$
and a coding operation on each node is given as linear operations
over input or incoming information, 
the code is called a {\it vector-linear code}.
The paper \cite{HOKC} showed the following lemma.

\section{Two kinds of formulations over single source and single terminal network model}\Label{S9}
To consider the relation between 
two kinds of formulations for our codes,
we consider an acyclic network with a single source and a single terminal,
where Eve is allowed to select an arbitrary  $r$-subset channels to access, 
which is called {\it $r$-wiretap network} \cite{Cai2002,CY,ACLY,LYC}.
We define the first capacity $C_1$ as the maximum transmission rate 
when random number generation is allowed in an arbitrary intermediate node.
We define the second capacity $C_2$ as the maximum transmission rate 
when random number generation is not allowed in an arbitrary intermediate node.
For our analysis on the capacities of the given network, we introduce two kinds of minimum cuts.
A node is called a pseudo source node when 
the node has only outgoing edges but has no original message to be transmitted.
Since a pseudo source node is not the source node nor the terminal node,
it is classified as an intermediate node.
The first kind of minimum cut $\mincut_1$ is the minimum number of edges crossing a line separating
the source and terminal nodes.
The second kind of minimum cut $\mincut_2$ is the minimum number of edges crossing a line separating
the source and terminal nodes 
when we remove all edges outgoing from pseudo source nodes.
Edges out-going from pseudo source nodes 
are counted in $\mincut_1$, but are not counted in $\mincut_2$.

\begin{lemma}[\protect{\cite[(35), (36), and Example 1]{CH}}]
For $r$-wiretap network,
we have 
\begin{align}
C_2=&
\mincut_2-r,\Label{E9-2} \\
\mincut_2-r\le
C_1 \le & 
\mincut_1-r.\Label{E9-2B}
\end{align}
\hfill $\square$\end{lemma}

As a special case analysis, we have the following corollary.
\begin{corollary}[\protect{\cite[Example 1]{CH}}]\Label{Cor}
When an $r$-wiretap network has no pseudo source node, 
we have 
\begin{align}
C_2=C_1=
\mincut_1-r.
\end{align}
\hfill $\square$\end{corollary}

The network given in Fig. \ref{6nodes} shows that
a network has different rates $\mincut_1$ and $\mincut_2$. 
This network has a linear code to realize $\mincut_1-r$ when $r=2$, 
which implies the equality of the second inequality in \eqref{E9-2B}.
Since $C_1=1$ and $C_2=0$, $C_1$ is strictly larger $C_2$.

\begin{figure}[htbp]
\begin{center}
\includegraphics[scale=0.7]{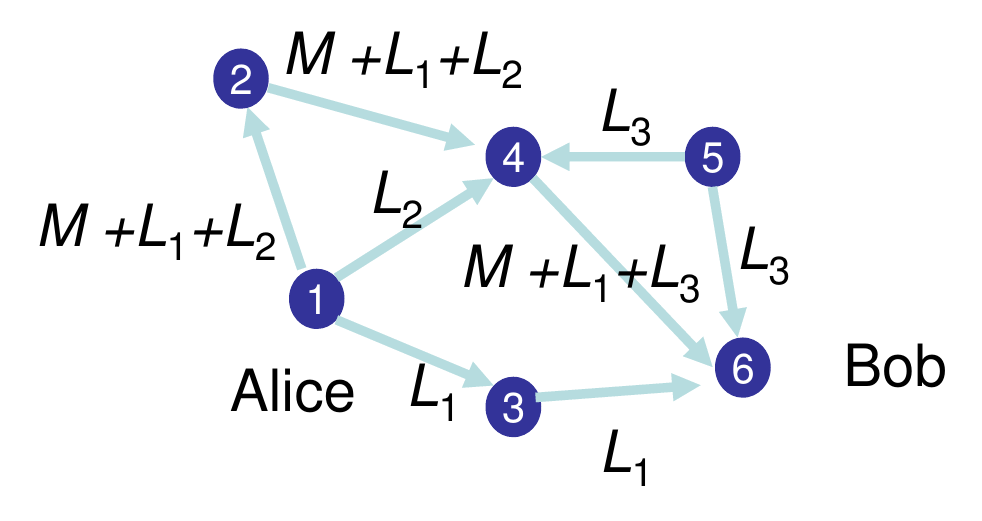}
\end{center}
\caption{$r$-wiretap network with equality of the second inequality in \eqref{E9-2B}.
Node 1 is the source node and Node 6 is the terminal node.
Node 5 is a pseudo source node.
Thus, $\mincut_2=2$ and $\mincut_1=3$.
When intermediate nodes are allowed to generate random variables,
the presented code achieves $\mincut_1-r$ when $r=2$.
The source node (Node 1) has the message $M$ and two scramble variables 
$L_1,L_2$.
The pseudo source node (Node 5) has another scramble variable $L_3$.
Even when Eve wiretaps arbitrary two edges, she cannot get 
arbitrary information for the message $M$.}
\Label{6nodes}
\end{figure}%

\section{Unconventional attack models}
\begin{figure}[htbp]
    \begin{tabular}{c}
      \begin{minipage}[t]{0.9\hsize}
        \centering
        \includegraphics[keepaspectratio, scale=0.6]{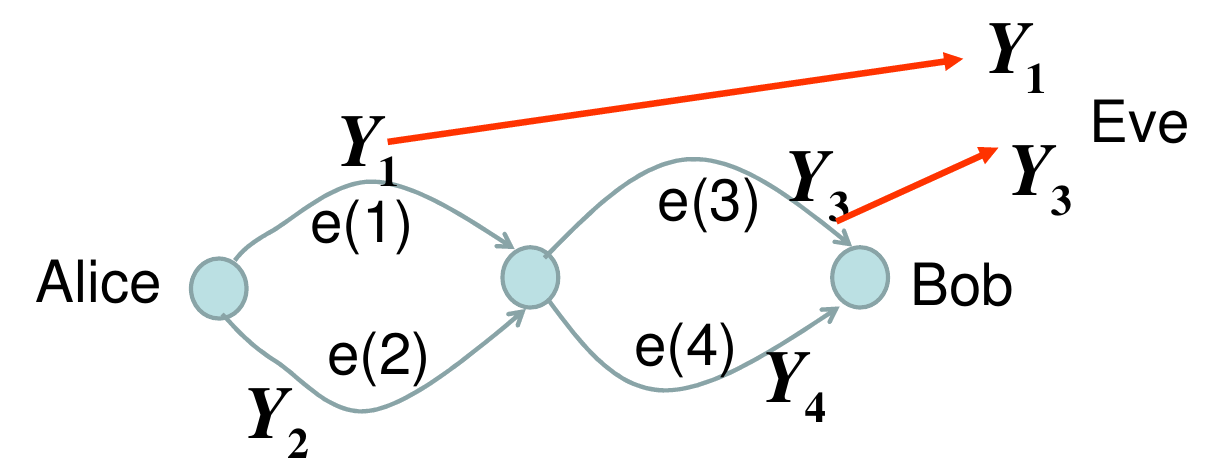}
        \caption{Deterministic and passive attack over one-hop relay network}
        \label{passive}
      \end{minipage} \\
      \begin{minipage}[t]{0.9\hsize}
        \centering
        \includegraphics[keepaspectratio, scale=0.6]{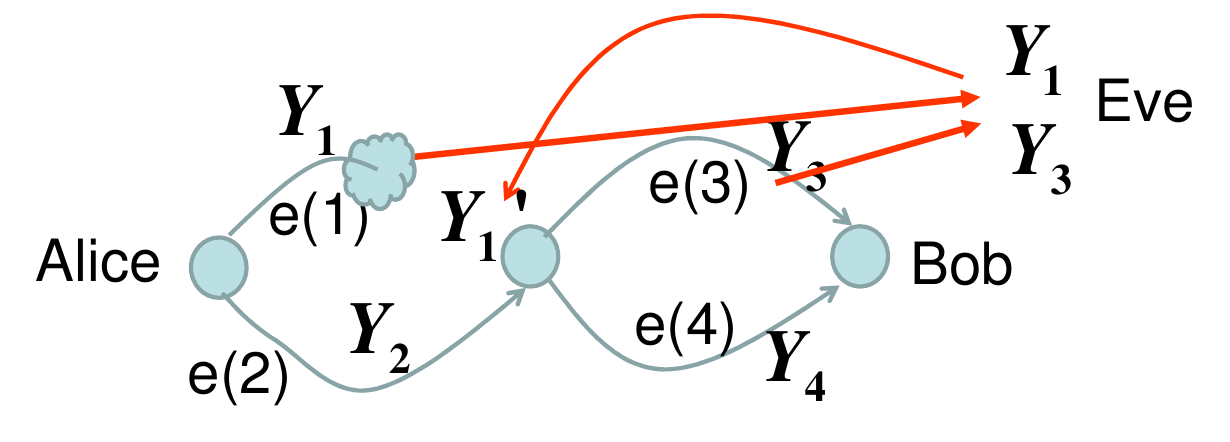}
        \caption{Active attack over one-hop relay network}
        \label{active}
      \end{minipage}\\
      \begin{minipage}[t]{0.9\hsize}
        \centering
        \includegraphics[keepaspectratio, scale=0.6]{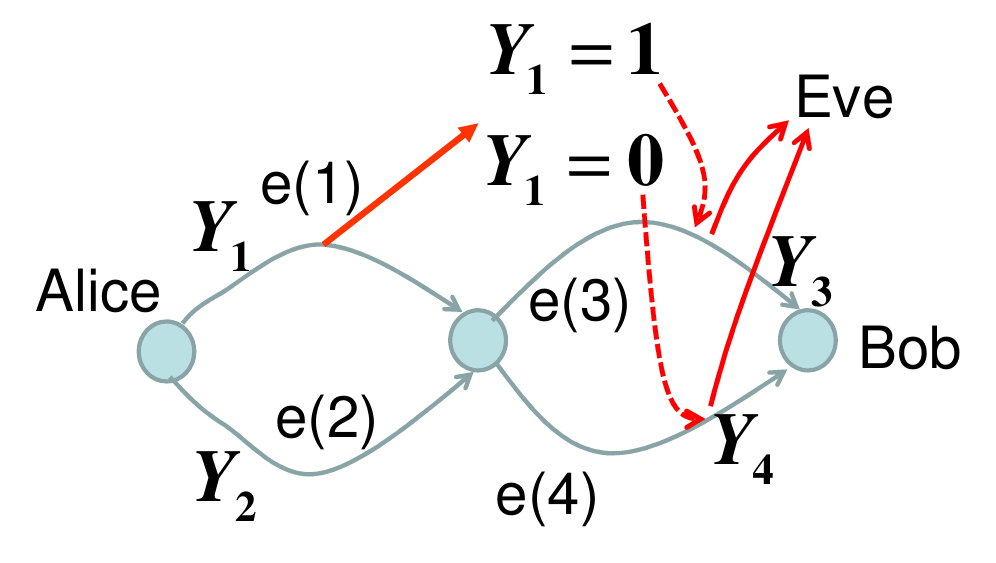}
        \caption{Adaptive attack over one-hop relay network}
        \label{adaptive}
      \end{minipage}
    \end{tabular}
  \end{figure}

\Label{SS}
This paper aims to discuss two unconventional types of attacks
when the network model has a multiple-layer structure, like
a one-hop relay network.
When a one-hop relay network is given, 
a conventional type of attack is given as Fig. \ref{passive}.
One is an {\it active attack} as Fig. \ref{active}.
In order to get more information,
the eavesdropper changes the information on attacked edges.
That is, after the above modification on the first attacked edge,
the eavesdropper, Eve, may have a possibility
to get more information via the second attack.
Such an attack is called an active attack.
If the attack is not an active attack, it is called a passive attack.

To discuss an active attack, we need to be careful for 
the causality and the time-ordering.
The paper \cite{HOKC} carefully handled this problem,
and showed the following theorem, which guarantees that
an arbitrary active attack cannot improve Eve's information
when the code is linear.

\begin{theorem}[\protect{\cite[Theorem 1]{HOKC}}]\Label{NMT}  
Suppose that coding operations on all nodes
are linear.
The information that Eve gets via an active attack on edges
can be simulated by Eve by using 
the information that Eve gets via a passive attack on the same edges.
\hfill $\square$\end{theorem}  

Although the paper \cite{Yao2014} addressed a part of active attacks,
its treatment is limited to a limited class of active attacks.
The paper \cite{HC21a} covers a more general class of active attacks.

The second is an {\it adaptive attack} as Fig. \ref{adaptive}.
In order to get more information,
Eve changes what edge is to be attacked
according the information on the previous attacks.
Such an attack is called an adaptive attack.
If the attack is not an adaptive attack, it is called a deterministic attack.
That is, a conventional attack is called 
a deterministic and passive attack.

\section{Problem over one-hop relay network}\Label{S3}
First, we consider a {\it one-hop relay network} as Fig. \ref{passive}.
This network has only one intermediate node in addition to 
the source node and the destination node.
The intermediate node connects with the source node via two edges, i.e., two channels,
$e(1)$ and $e(2)$.
Also, the intermediate node connects with the destination node via two edges, i.e., two channels,
$e(3)$ and $e(4)$.
The aim of this network is the following.
Alice intends to her message from the source node to 
Bob at the destination node. 
Here, we consider two types of attacks by an adversary, Eve.
In the passive attack, Eve can eavesdrop on two edges in total,
i.e., she can eavesdrop on one edge among $e(1)$ and $e(2)$,
and can another edge among $e(3)$ and $e(4)$.
In the active attack, Eve is allowed to change it to get more information.

Here, for the secrecy, we adopt imperfect secrecy as follows.
When $M$ is the message and
Eve's information $Z_E$ satisfies the condition $I(M;Z_E) =0$ 
for all of Eve's possible attacks, 
the code is called {\it perfectly secret}.
When the condition $I(M;Z_E) < \log d$ holds for all of Eve's possible attacks, 
the code is called {\it imperfectly secret} \cite{Bhattad} (or weakly secret). 
Otherwise, it is called {\it insecure}. 
In other words,
our code is imperfectly secret,
when there exists no function $\tilde{\psi}$ such that $\tilde{\psi}(Z_E) = M$.
Also, we say that the code is perfectly secret,
when $Z_E$ is Eve's information and $I(M;Z_E) = 0$ for all Eve's possible
attacks.

We choose $L \in\ZZ_d$ as the uniform scramble random variable
generated at the source node. Suppose that the intermediate
node generates another uniform scramble random variable
$L'\in \ZZ_p$. 
We introduce a scalar-linear code as follows.
\begin{align}
Y_1 := L, \quad Y_2 := M + L. \Label{Eq3B}
\end{align}
Then, the intermediate node applies the code $\varphi$ as
\begin{align}
{Y}_3:= L',\quad
{Y}_4:=Y_2-Y_1+L'. \Label{Eq2F}
\end{align}
The decoder $\psi$ is written as $\psi({Y}_3,{Y}_4):={Y}_4-{Y}_3$,
which equals $Y_2-Y_1+L'-L'=M+L-L +L'-L'=M$.
This code satisfies the correctness.
The pair $(Y_1,Y_3)$ is independent of $M$.
Similarly, the pairs $(Y_1,Y_4)$, $(Y_2,Y_3)$, and $(Y_2,Y_4)$ are independent of $M$.
Thus, this code is perfectly secret for passive attacks, and has the transmission rate $\log d$.

Even when Eve selects the attacked edge $e(3)$ or $e(4) $ based on the 
observation on the first attack,
Eve gets no information for the message $M$.
Hence,
this code is secret even for an adaptive attack.
Due to the linearity, even when 
Eve changes the information on the first attacked edge,
Eve cannot improve her performance due to Theorem \ref{NMT}.
Thus, this code is secret even for an adaptive and active attack.

However, when the intermediate node cannot generate 
a uniform scramble random variable,
no scalar-linear code is imperfectly secret 
over a quotient ring $\ZZ_d$ for passive attacks, as shown in \cite[Theorem 1]{HC21b}.
In other words, for an arbitrary scalar-linear code over 
a quotient ring $\ZZ_d$,
Eve has a passive attack to recover the message $M$ perfectly.

\section{Non-linear codes over one-hop relay network}\Label{S4}
To realize imperfect secrecy without random number generation on 
the intermediate node,
we introduce the following non-linear code.
For simplicity, it is assumed that Alice transmits only the message $M \in \ZZ_d$ and an arbitrary edge can transmit only one binary information.
We consider the following code, which requires 
the binary uniform scramble random variable $L\in \ZZ_d$
at the source node.

The encoder $\phi$ is given in the same way as \eqref{Eq3B}.
Then, our non-linear code $\varphi$ in the intermediate node is given as
\begin{align}
{Y}_3:= {Y}_1({Y}_2-{Y}_1) , \quad
{Y}_4:=({Y}_1+1)({Y}_2-{Y}_1). \Label{Eq2}
\end{align}
The decoder $\psi$ is given as $\psi({Y}_3,{Y}_4):={Y}_3+{Y}_4$.
In this code, ${Y}_3$ and ${Y}_4$ have the following form;
\begin{align}
{Y}_3= LM, \quad
{Y}_4=LM+M.
\end{align}
Thus, the decoder recovers $M$ as $Y_4-Y_3$
nevertheless the value of $L$.
We call the above code the {\it standard non-linear code}.

Consider the case when Eve eavesdrops $Y_1=L$ in the first step.
Suppose that Eve eavesdrops $Y_3$ in the second step.
When $L$ is invertible, she can recover $M$ by $Y_1^{-1}Y_3 $.
But, when $L$ is not invertible, she cannot recover $M$ 
because the map $M \mapsto LM $ is not one-to-one.
Suppose that Eve eavesdrops $Y_4$ in the second step
When $L+1$ is invertible, she can recover $M$ by $(Y_1+1)^{-1}Y_4 $.
But, when $L+1$ is not invertible, she cannot recover $M$ because the map $M \mapsto (L+1)M $ is not one-to-one.

Consider the case when Eve eavesdrops $Y_2=M+L$ in the first step.
Suppose that Eve eavesdrops $Y_3$ in the second step.
To recover $M$, she needs to find $M$ to satisfy $Y_3=Y_2 M-M^2 $.
However, the map $M \mapsto Y_2 M-M^2$
is not always one-to-one.
For example, 
when $d=2$ and $Y_2=1$, $Y_2 M-M^2$ is always zero.
Also,
when $d>2$ and $Y_2=0$, $Y_2 M-M^2$ has the same value with
$M=1,-1$.
Thus, Eve cannot recover the original information $M$.
Suppose that Eve eavesdrops $Y_4$ in the second step.
To recover $M$, she needs to find $M$ to satisfy $Y_4=Y_2 M+M-M^2 $.
Similarly, the map $M \mapsto Y_2 M+M-M^2$
is not always one-to-one.
For example, 
when $d=2$ and $Y_2=0$, $Y_2 M+M-M^2$ is always zero.
Also,
when $d>2$ and $Y_2=-1$, $Y_2 M+M-M^2$ has the same value with
$M=1,-1$.
In this way,
we find that this code is imperfectly secret
for deterministic and passive attacks.
That is, there exists a secret code over deterministic and passive attacks.

When $d=2$, 
The reference \cite[Appendix]{HC21b} calculated
the mutual information, i.e., the amount of information leakage
as follows.
\begin{align}
  I(M;{Y}_1,{Y}_3)= I(M;{Y}_1,{Y}_4)
  = I(M;{Y}_2,{Y}_3)=  I(M;{Y}_2,{Y}_4)=\frac{1}{2},\Label{E9}
\end{align}

Further, as shown in Lemma \ref{T6} with $d=2$, 
when 
Eve cannot recover the message $M$ perfectly with 
an arbitrary deterministic and passive attack in the code,
the network code is limited to the standard non-linear code or a code equivalent to the standard non-linear code.

\begin{lemma}[\protect{\cite[Lemma 1]{HC21b}}]\Label{T6}
Suppose that $d=2$ and 
a code $(\phi,\varphi,\psi)$ satisfies the following conditions.
Let $Y_1$ and $Y_2$ be the random variables generated by the encoder 
$\phi$ when $M$ is subject to the uniform distribution.
We assume that the random variables $(Y_3,Y_4):=\varphi (Y_1,Y_2)$
satisfies the following conditions.
\begin{description}
\item[(C1)] The relation $\psi (Y_3,Y_4)=M$ holds.
\item[(C2)] There is no deterministic function $\tilde{\psi}$
from $\ZZ_2^2$ to $\ZZ_2$ satisfying one of the following conditions.
\begin{align}
\tilde{\psi}(Y_1,Y_3)=M,\quad
\tilde{\psi}(Y_1,Y_4)=M, \\
\tilde{\psi}(Y_2,Y_3)=M,\quad
\tilde{\psi}(Y_2,Y_4)=M.
\end{align}
\end{description}
Then, there exist single-input functions
$f_1,f_2,f_3,f_4,f_5$ on $\ZZ_2$ such that
$\bar{Y}_i:=f_i(Y_i)$
and $\bar{M}:=f_5(M)$ satisfy \eqref{Eq2} and \eqref{Eq3B}
with a scramble random variable $L$ while the variable $L$ might be correlated with $M$.
\hfill $\square$\end{lemma}

However, the above theorem does not holds
when $d >2$.
In this case, the paper \cite{HC21b} proposed another non-linear code.
For this aim, the paper \cite{HC21b} introduced an 
anti-Latin square. A matrix $(a_{i,j})$ on $\ZZ_d$ is called an 
{\it anti-Latin square} 
when each row and each column have duplicate elements, 
which is the opposite requirement to a Latin square. 
For example, the following shows
$3\times 3$ and $4\times 4$ anti-Latin square matrices.
\begin{align}
\left(
\begin{array}{ccc}
0 & 1 & 0 \\
1 & 1 & 2 \\
0 & 2 & 2
\end{array}
\right), \quad
\left(
\begin{array}{ccc}
0 & 2 & 2 \\
0 & 1 & 0 \\
1 & 1 & 2
\end{array}
\right), \quad 
\left(
\begin{array}{cccc}
0 & 1 & 3 & 3 \\
0 & 1 & 2 & 0 \\
1 & 1 & 2 & 3 \\
0 & 2 & 2 & 3 
\end{array}
\right), \quad
\left(
\begin{array}{cccc}
0 & 0 & 1 & 0 \\
1 & 1 & 1 & 2 \\
3 & 2 & 2 & 2 \\
3 & 0 & 3 & 3 
\end{array}
\right).
\label{e21} 
\end{align}

Using a pair of 
$d\times d$ anti-Latin square matrices $(a_{i,j})$ and $(b_{i,j})$,
the paper \cite{HC21b} proposed a non-linear code as follows.
The encoder $\phi$ is given in the same way as \eqref{Eq3B}.
Then, our non-linear code $\varphi$ in the intermediate node is given as
\begin{align}
{Y}_3:= a_{Y_1,Y_2} , \quad
{Y}_4:=b_{Y_1,Y_2} . \Label{Eq2T}
\end{align}
We call the above code the {\it anti-Latin code} of the pair of
anti-Latin square matrices $(a_{i,j})$ and $(b_{i,j})$.

For example,  when the pair $(a_{i,j})$ and $(b_{i,j})$ is chosen by 
the first and second matrices (or the 
the third and fourth matrices) in \eqref{e21}, 
the map $(Y_1,Y_2)\mapsto (a_{Y_1,Y_2},b_{Y_1,Y_2})$
is one-to-one. So, the anti-Latin code is uniquely decodable.
To see the equivalent condition to the unique decodability,
we define the set
\begin{align}
\Xi_{z,m}(\varphi^{(3)},\varphi^{(4)}):=
\varphi^{(4)} (\{(l,l+m)| \varphi^{(3)}(l,l+m)=z\})
 \hbox{ for }z,m
\in \ZZ_d.
\end{align}
When $Y_4=z$ and $M=m$,
the set
$\Xi_{z,m}(\varphi^{(3)},\varphi^{(4)})$ expresses 
the possible values of $Y_4$.
Hence, the unique decodability is equivalent to 
the relation 
\begin{align}
\Xi_{z,m}(\varphi^{(3)},\varphi^{(4)})
\cap
\Xi_{z,m'}(\varphi^{(3)},\varphi^{(4)})
=\emptyset \hbox{ for }m\neq m'\in \ZZ_d,~z \in \ZZ_d.
\label{BNU}
\end{align}
Although there is no decodable anti-Latin code for $d=2$,
a decodable anti-Latin code exists for $d>2$ as follows.

\begin{lemma}[\protect{\cite[Section V]{HC21b}}]\Label{TT}
For $d>2$, there exists a pair of 
$d\times d$ anti-Latin square matrices $(a_{i,j})$ and $(b_{i,j})$
such that its anti-Latin code is decodable.
\hfill $\square$\end{lemma}


\section{Active attack to non-linear code over one-hop relay network}\Label{S6}
Next, we consider active attacks over a one-hop relay network, as Fig. \ref{active}.
Eve has the following two active attacks for the standard 
non-linear code. 
\begin{description}
\item[(i)]
When Eve attacks on $e(1), e(3)$
and replaces $Y_1$ by $1$, we have 
$Y_3+1-Y_1=M$.

\item[(ii)]
When Eve attacks on $e(1), e(4)$
and replaces $Y_1$ by $0$, we have 
$Y_4-Y_1=M$.
\end{description}
Therefore, the code presented in Section \ref{S5}
is insecure under active attacks.

Since Lemma \ref{T6} guarantees the non-existence of another 
imperfectly secret code over deterministic and passive attacks for $d=2$,
the above discussion shows the non-existence of an imperfectly secret code over active attacks for $d=2$.

However, Lemma \ref{T6} does not hold for $d\neq 2$.
In fact, any anti-Latin code is imperfectly secret even under 
active attacks as follows.
Assume that Eve eavesdrops $e(1)$ and $e(3)$.
Even when Eve replaces $Y_1$ by any value,
any row vector of $a_{i,j}$ has duplicate elements.
Hence, Eve has a possibility that she cannot recover the message
from $Y_1$ and $Y_3$.
The same discussion can be applied to other pairs of 
edges eavesdropped by Eve.

\section{Adaptive attack to non-linear code over one-hop relay network}\Label{S5}
Next, we discuss adaptive attacks over one-hop relay network
even when active modification is not allowed.
Eve has the following adaptive attack to recover the message $M$, as Fig. \ref{adaptive}.

\begin{description}
\item[(i)]
First, Eve eavesdrops $e(1)$.
When ${Y}_1=1$, 
she eavesdrops $e(3)$, and recovers $M$ as 
${Y}_3+1={Y}_2-1+1= M$.
When ${Y}_1=0$, she eavesdrops $e(4)$, and
 recovers $M$ as ${Y}_4={Y}_2=M$.
\end{description}

Thus, the non-linear code given in Section \ref{S4}
is not imperfectly secret for adaptive attacks.
Since Lemma \ref{T6} guarantees the non-existence of another 
imperfectly secret code over deterministic and passive attacks for $d=2$,
there exists no imperfectly secure code over adaptive attacks
in this network model with $d=2$.
This fact shows that an adaptive attack is powerful for this kind of non-linear code with $d=2$
even when it has no active modification.
In fact, when $d=2$, 
Eve also has the following adaptive attack to recover the message $M$. 
\begin{description}
\item[(ii)]
First, Eve eavesdrops $e(2)$.
When ${Y}_2=1$, 
she eavesdrops $e(4)$ and
recovers $M$ as ${Y}_4={Y}_1+1=M$.
When ${Y}_2=0$, she eavesdrops $e(3)$ and recovers $M$ as ${Y}_3={Y}_1=M$.
\end{description}

However, Lemma \ref{T6} does not hold for $d\neq 2$.
In fact, 
any anti-Latin code is imperfectly secret even under 
adaptive and active attacks as follows.
Assume that Eve eavesdrops $e(1)$ at the first step. 
Even when Eve replaces $Y_1$ by any value,
any row vector of $(a_{i,j})$ 
and any row vector of $(b_{i,j})$ 
have duplicate elements.
When Eve observes such duplicate elements in the second observation,
she cannot recover the message $M$.
Hence, 
whichever $e(3)$ or $e(4)$ Eve eavesdrops in the second step,
she has a possibility that she cannot recover the message
$M$ from her observed information.
The same discussion can be applied to 
the case when Eve eavesdrops $e(2)$
by considering 
column vectors of $(a_{i,j})$ and $(b_{i,j})$. 
Therefore, there exists an imperfectly secret code for 
under adaptive and active attacks when $d>2$.

The discussion of this section is based on the paper \cite{CH}.
But, it did not discuss adaptive attacks 
for anti-Latin codes.

\section{Vector-linear codes over one-hop relay network}\Label{S7}
To resolve this problem, 
we employ a vector-linear code.
A vector-linear code can be realized 
when the network of Fig. \ref{passive} is used at least twice.
Suppose that the source node generates three uniform scramble random variables $L_1,L_2,L_3\in \ZZ_d$.
Using these variables, we define a vector-linear code as follows.
The first transmission sends the following;
\begin{align}
{Y}_1:=L_1, \quad {Y}_2:=M+L_1, \Label{Eq3C}
\end{align}
and the second transmission sends the following;
\begin{align}
{Y}_1':=L_2, \quad {Y}_2':=L_3+L_2, \Label{Eq3D}
\end{align}
Then, the intermediate node applies the code $\varphi$ as
\begin{align}
{Y}_3:= Y_2'-Y_1',\quad
{Y}_4:=Y_2-Y_1+Y_2'-Y_1'. \Label{Eq2G}
\end{align}
In this code, nothing is transmitted in the second layer at the second transmission.
The decoder $\psi$ is given as $\psi({Y}_3,{Y}_4):={Y}_4-{Y}_3$,
which equals $Y_2-Y_1+Y_2'-Y_1'-(Y_2'-Y_1')=M+L_1-L_1=M$.
Then, the pair $(Y_1,Y_3)$ is independent of $M$.
Similarly, the pairs $(Y_1,Y_4)$, $(Y_2,Y_3)$, and $(Y_2,Y_4)$ are independent of $M$.
Thus, this code is perfectly secret for deterministic attacks, 
and has the transmission rate $\frac{1}{2}\log d$.
In the same reason as the above,
this code is secret for adaptive attack.
Then, due to the linearity, it is secure even for active and adaptive attacks.

The discussion until this section is summarized as Table \ref{non-linear}.

\begin{table}[htpb]
  \caption{Summary for one hop relay network 
  (Figs. \ref{passive}-\ref{adaptive}) with single shot setting}
\Label{non-linear}
\begin{center}
  \begin{tabular}{|l|c|c|c|} 
\hline
\multirow{3}{*}{Code} 
& deterministic & \multirow{2}{*}{active} &\multirow{2}{*}{adaptive} \\
& and passive & \multirow{2}{*}{attack}& \multirow{2}{*}{attack}\\
& attack & &\\
\hline
\hline
\multirow{2}{*}{scalar-linear code over $\ZZ_d$} & 
\multirow{2}{*}{insecure}& 
\multirow{2}{*}{insecure}&
\multirow{2}{*}{insecure}
\\
& &&\\
\hline
\multirow{2}{*}{scalar-non-linear code over $\ZZ_2$} & imperfectly  & \multirow{2}{*}{insecure} &\multirow{2}{*}{insecure}\\
& secret& &\\
\hline
scalar-non-linear code  & imperfectly & imperfectly &imperfectly\\
over $\ZZ_d$ with $d>2$ & secret & secret &secret\\
\hline
\multirow{2}{*}{vector-linear code over $\ZZ_d$} & perfectly & perfectly &perfectly \\
& secret & secret &secret\\
\hline
  \end{tabular}
\end{center}
\end{table}

\section{Non-local linear codes over unicast relay network}\Label{S8}
\subsection{Capacity formula}
The above type of codes can be generalized over 
unicast relay network.
Before the detailed discussions, we remark on the relation between two kinds of linear codes
over multiple uses of a given network.
When coding operations on intermediate nodes are fixed to linear operations,
such a model is called a linear network model.
In this case, we have the freedom to design only 
coding operations in the source and destination nodes.
Such coding operations are called an encoder and a decoder.
Thus, operations at intermediate nodes are not designed.
Such a code is called a {\it local code}, and was studied in the papers \cite{CH}.
However, the code presented in Section \ref{S6} 
is not classified to this class of codes.
In contrast, when 
all operations at intermediate nodes are designed to achieve a higher transmission rate,
such a code is called a {\it non-local code}.

\begin{figure}[htbp]
\begin{center}
\includegraphics[scale=0.49]{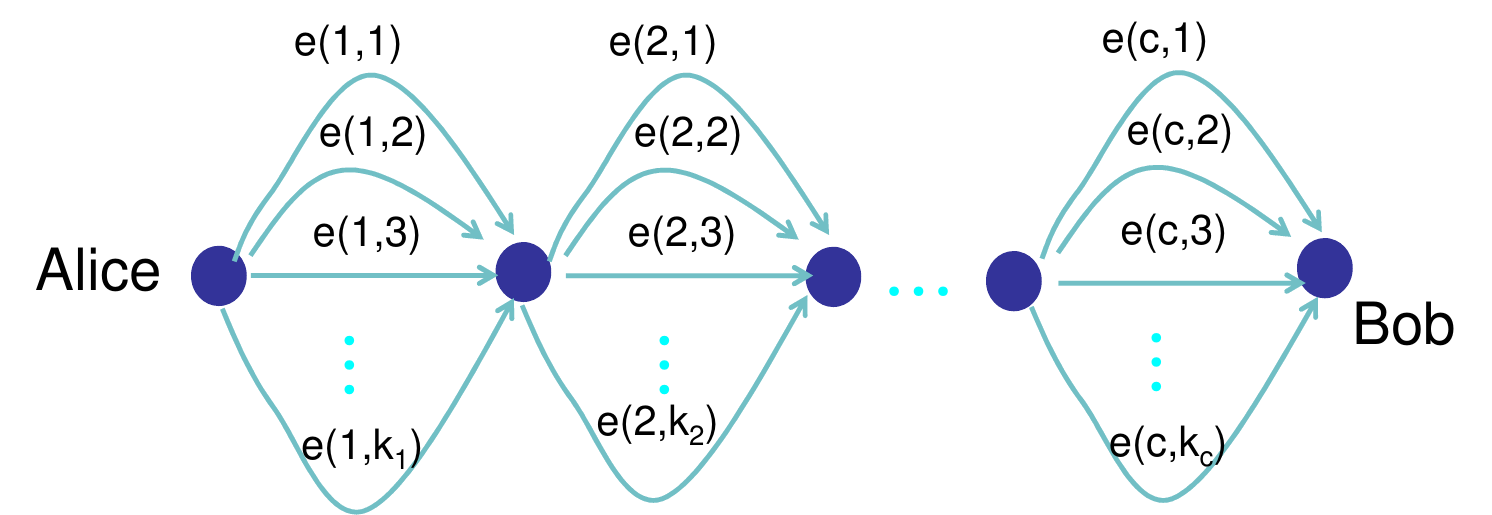}
\end{center}
\caption{Unicast relay network}
\Label{unicast}
\end{figure}%

In the following, we consider 
non-local codes over a {\it unicast relay network} given as Fig. \ref{unicast}.
But, for simple discussion, we assume that
each edge sends one element of a finite field $\FF_q$ 
instead of an element of $\ZZ_d$
for a single channel use.
A unicast relay network is composed 
of the following group structure of the intermediate nodes.
This network has 
$c-1$ intermediate nodes, which are labeled from 
the first intermediate node to the $c-1$-th intermediate node.
It has only 
one source node and one terminal node, which are regarded as
the $0$-th node and the $c$-th node, respectively.
Each intermediate node has incoming edges and outgoing edges.
For $i=1, \ldots, c$,
this network has $k_i$ edges between the $i-1$ and $i$-th nodes. 
At one channel use, each edge $e(i,j)$ transmits the information ${Y}_{i,j}$ for $i=1,\ldots, c$ 
and $j=1, \ldots, k_i$ that takes values on $\{1, \ldots, d\}$.

In this model, it is assumed that Eve eavesdrops $r_i$ edges 
$\vec{Y}_{i,s_i}:=({Y}_{i, s_i(1)}, \ldots, {Y}_{i, s_i(r_i)}) $
among $k_i$ edges 
$\overline{Y}_{i}:=({Y}_{i, j})_{j=1, \ldots, k_i} $
between the $i-1$ and $i$-th nodes.
Under this notation, 
the function $s_i$ expresses the edges eavesdropped by Eve.
In other words, she can eavesdrop $\sum_{i=1}^c r_i$ edges totally.
Eve is allowed to make 
adaptive and active attacks, which are 
stronger attacks than conventional attacks.

The first capacity $C_1$ is defined as the maximum transmission rate 
when random number generation is allowed in an arbitrary intermediate node.
The second capacity $C_2$ is defined as the maximum transmission rate 
when random number generation is not allowed in an arbitrary intermediate node.
Then, we have the following capacity theorem.
\begin{theorem}[\protect{\cite[Theorem 3]{CH}}]
\Label{TT3}
We have the capacity formulas
\begin{align}
C_1 =& 
\log q \min_{1 \le j \le c} (k_j-r_j) \Label{Cap-1},
\\
C_2=&
\log q \min_{1 \le j \le c}
(k_j-r_j) 
\frac{(k_{j+1}-r_{j+1}) \cdots (k_c- r_c) }{k_{j+1} \cdots k_c }.\Label{Cap-2}
\end{align}
Further, these capacities can be attained by non-local linear codes
under the respective classes of non-local codes.
\hfill $\square$\end{theorem}

Further, the paper \cite{CH} showed that 
these capacities cannot be improved
even when the attack is limited to deterministic and passive attacks.
This fact shows that
adaptive and active attacks cannot improve Eve's ability in this network
when we use the proposed code.

Here, we discuss the relation to existing results with respect to the difference between two capacities $C_1$
and $C_2$.
A larger part of preceding studies discuss 
the capacity $C_1$, i.e.,
the capacity (or capacity region) with no restriction of
randomness generated at intermediate nodes.
For example, 
as stated in Corollary \ref{Cor},
the equation $C_1=C_2$ holds in $r$-wiretap network
when $\mincut_1=\mincut_2$.
However, the paper \cite{CY-07} showed an example for network
such that the capacity can be improved by
randomness generated at intermediate nodes.
In the example, only one edge connects the source node.
This example seems natural because 
usually, the secure transmission can be done by use of the 
difference among information on different edges connected to the same node.

The papers \cite{CHK-10,CHK-13} addressed the difference between the existence and non-existence of 
randomness generated at intermediate nodes in another network
only for deterministic attacks.
But, these papers did not derive the capacities $C_{1}$ and $C_{2}$ exactly.
because their analysis depends on special codes.
Therefore, the analysis in \cite{CH}
can be considered as the first derivation of the difference between 
the capacities $C_{1}$ and $C_{2}$ except for the case when 
only one edge connects the source node.

\subsection{Ideas for proof}
Since the proof of this theorem is too complicated, we omit the proof.
Here, we simply explain what technical lemmas are employed for the proof.
The converse part was shown by using the following lemma.

We denote the set $\{1, \ldots, k\}$ by $[k]$,
and denote the collection of subsets $S\subset [k]$ with cardinality 
$r$ by ${ [k] \choose r}$.
Now, we consider the random variables
$X,\vec{Y}_1, \ldots, \vec{Y}_k$.
For an arbitrary subset $S \subset [k]$, we denote the tuple of random variables
$(\vec{Y}_{s})_{s \in S}$ by $\vec{Y}_S$.
We can show the following two lemmas.

\begin{lemma}[\protect{\cite[(59)']{CH-cor}}]\Label{LH10}
We have
\begin{align}
\sum_{S \in { [k] \choose r}}
H(\vec{Y}_{S}| X)
\ge &
{ k-1 \choose r-1}
H(\vec{Y}_{[k]}| X)
=
\frac{r}{k}{ k \choose k-r}
H(\vec{Y}_{[k]}| X).\label{ZBX}
\end{align}
\hfill $\square$\end{lemma}

Originally, the paper \cite{CH} stated 
a statement different from the above
as Lemma 4. 
However, its derivation is not correct and was not used for the converse proof,
as explained in \cite{CH-cor}.
Lemma \ref{LH10} is the correct version, and 
is used in Eq. (47) in the converse part in \cite{CH}. 
Although Lemma \ref{LH10} has been already known as Han's inequality \cite{Han},
it can be shown by Lemma \ref{Lemma-1} shown in \cite{CH},
in the following way.

Here, we show Lemma \ref{LH10} by using Lemma \ref{Lemma-1}.
Any element $a \in [k]$ is contained in exactly ${k-1 \choose r-1}  $ members of $ {[k] \choose r}$.
So, we apply Lemma \ref{Lemma-1} 
to the case with ${\cal S}_h= {[k] \choose r}$ 
and $h={k-1 \choose r-1} $.
Thus, we have Eq. \eqref{ZBX}.

\begin{lemma}[\protect{\cite[Lemma 5]{CH}}] \Label{Lemma-1}
Let ${\cal S}_h$ be a collection of subsets of $[k]$.
When an arbitrary element of $[k]$ is contained in exactly $h$ members of ${\cal S}_h$,
we have
\begin{equation} \Label{eq:Lemma-11}
\sum_{S \in {\cal S}_h} H(\vec{Y}_S|X) \ge h H(\vec{Y}_{[k]}|X).
\end{equation}
\hfill $\square$\end{lemma}

In contrast,
the direct part was shown by using the following technical lemma.

\begin{lemma}
\Label{L425}
For an arbitrary prime power $q$, arbitrary two natural numbers $k>r$,
there exist a natural integer $n_{k,r}$ 
and $r$ vectors $v_1, \ldots, v_r \in \ZZ_{q^{n_{k,r}}}^{k}$
such that $v_{i,j}=\delta_{i,j} $ for $j=1, \ldots,m$ and 
the $r \times r$ matrix $(v_{i,s(j)})_{i,j}$ is invertible
for an arbitrary injective function $s$ from $\{1, \ldots, r\}$ to 
$\{1, \ldots, k\}$.
\hfill $\square$\end{lemma}

Lemma \ref{L425} has been shown in various topics, 
the wiretap channel II introduced by Ozarow and Wyner \cite{Ozarow} and
secret sharing \cite{Shamir,Blakley2}, etc.
In the wiretap channel II model,
a secret message is encoded to a codeword in an $n_{k,r}$-length code. 
A wiretapper can eavesdrop on arbitrary $r$ components 
out of $k$ parallel channels, but may have no information about the message. 
A linear code, e.g., a Reed-Solomon code can serve as the code. 
It is called a $(k,r)$ code for wiretap channel II,
and satisfies the condition for Lemma \ref{L425}.
Also, such a code is called a 
maximum distance separable (MDS) code \cite{Blakley}. 
This lemma also can be regarded as a very simple and special case of the code in \cite[Section III]{CY}.

Since the network coding with local codes \cite{Yao2014,HC21a}
includes the wiretap channel II model
as the case with parallel channels \cite{Zhang},
it can be considered that the analysis \cite{Yao2014,HC21a}
contains a more general statement than Lemma \ref{L425}.

Recently, this lemma was generalized to its symplectic version as Lemma 2 of \cite{Song}.
The symplectic version is useful for analyzing the quantum version of 
private information retrieval.
Also, the recent paper \cite{HS23}
generalized this lemma to various ways
by using the concept of 
multi-target monotone span program (MMSP)
to discuss various types of quantum extension of 
MDS codes.

\section{Conclusion}\Label{S11}
We have reviewed several results under adaptive and active attacks
over a one-hop relay network and unicast relay network.
Also, we have reviewed the recent result for 
the difference between 
 the existence and the non-existence
of random number at intermediate nodes.
These results show how to handle such unconventional attacks.
Although the papers \cite{HC21b,CH} did not discuss
the imperfect secrecy for 
an anti-Latin code under adaptive attacks,
we have proved it in this paper.
The paper \cite{CH} also generalized the result presented in Section \ref{S8}
to a homogeneous multicast rely network.
Although the review of this extension is omitted,
the paper \cite{CH} presented a network code that works even with 
adaptive and active attack in these networks.
However, the analysis on such unconventional attacks over a network
is still limited.
Further, it is desired to develop network codes for  
such unconventional attacks.

The paper \cite{HC21b} proposed the concept of 
an anti-Latin square, and no other paper discussed it.
Based on Latin squares, the famous puzzle, Sudoku,
has been invented and it has been enjoyed among many people.
Since generating a pair of anti-Latin squares
is not so trivial and is an interesting combinatorics problem,
it can be expected to be developed into another puzzle.
Also, it is expected that 
a more complicated combination of anti-Latin squares
generates useful non-linear codes for more complicated network models.
Although the decodability of an anti-Latin code
is equivalent to \eqref{BNU},
it is not clear whether 
this condition is equivalent to
the condition that the map $(Y_1,Y_2)\mapsto (a_{Y_1,Y_2},b_{Y_1,Y_2})$
is one-to-one. 
That is, while the latter condition is its sufficient condition,
its necessity is not clear.
This problem is also an interesting open problem.

Finally, we propose other open problems.
A subset $S$ of the set ${\cal A}(d)$ of $d\times d$ anti-Latin matrices
is called mutual decodable
when any two elements of $S$ form a decodable anti-Latin code.
A subset $S\subset {\cal A}(d)$ 
is called mutual one-to-one
when the map $(Y_1,Y_2)\mapsto (a_{Y_1,Y_2},b_{Y_1,Y_2})$
is one-to-one for any two elements $(a_{i,j})$ and $(b_{i,j})$ of $S$.
Then, we define two numbers $A(d)$ and $B(d)$ as
\begin{align}
A(d)&:= \max_{S: \subset {\cal A}(d)}\{
|S|: S\hbox{ is mutual decodable.} \} \\
B(d)&:= \max_{S: \subset {\cal A}(d)}\{
|S|: S\hbox{ is mutual one-to-one.} \}
\end{align}
The calculations of these values are another interesting open problem.
For Latin squares, a similar problem has been studied as follows.
Two Latin squares $(a_{i,j})$ and $(b_{i,j})$ is called orthogonal
when the map $(i,j)\mapsto (a_{i,j},b_{i,j})$ is one-to-one.
It is known that the maximum size of a set of $n \times n$ mutually orthogonal Latin squares is $n-1$ 
when $n$ is a prime or prime power
\cite{MacNeish,Mann}.
The general case of this maximum number has been actively studied in the area of combinatorics \cite{Bose}.

Indeed, a mutual decodable subset 
$S\subset {\cal A}(d)$ can be used to modify 
the one-hop relay network as follows.
That is, the calculation of $A(d)$ has the following operational meaning.
We modify the one-hop relay network by putting
$|S|$ channels in the second step,
and the code in the second step is defined by 
$|S|$ anti-Latin matrices.
In this case, we assume the following.
Eve is allowed to eavesdrop one edge in the first group and 
another edge in the second group, adaptively.
The receiver accesses only two edges in the second group
and the two edges are selected randomly.
In this situation, 
the above code based on a pairwise-decodable subset $S$ 
is imperfectly secret and decodable.

\end{document}